\documentclass{elsart}

\usepackage{natbib,psfig}

\usepackage{amssymb}

\begin{document}

\begin{frontmatter}



\title{AGB Stars: Summary and Warning}


\author{John C Lattanzio}

\address{School of Mathematical Sciences, PO Box 28M, 
Monash University, 3800, Australia}
\ead{j.lattanzio@sci.monash.edu.au}
\ead[url]{www.maths.monash.edu.au/$\sim$johnl/}
\begin{abstract}
% Text of abstract
We review the evolution and 
nucleosynthesis of AGB stars. We then discuss some of the
contentious issues and quantitative uncertainties in current models.
\end{abstract}


\end{frontmatter}

% 
%Local definitions are here
\def\msun{M_\odot}

% main text
\section{Introduction}
\label{intro}
The Asymptotic Giant Branch (AGB) phase is very short
but its importance is seen in its nucleosynthesis.
A revolution in stellar modelling
has taken place in the last 20 years, inspired jointly by this rich
nucleosynthesis and partly by new data. For example,
the isotopic data coming from pre-solar grains 
(see this volume) forces theorists to include species that
were previously ignored, species which are energetically of no
importance (i.e. they play no role in determining the stellar structure)
but which can be used to constrain the models.
Nucleosynthesis is now important as a tracer of temperature and mixing,
and not simply a by-product of energy generation.
But along with these advances come more quantitative demands.
It is now increasingly important to know what is known well and what is less sure. This is the goal of this paper.

\section{AGB Evolution and Nucleosynthesis}
\label{evoln}

All stars between about 1--8 $\msun$ will pass through the AGB phase.
A star will begin its AGB phase following core helium exhaustion. It is 
worth emphasizing again the tiny size of the core (essentially a white dwarf) 
in comparison to the enormous envelope, which extends over a few hundred 
solar radii. In more familiar terms, if the entire AGB star envelope were 
reduced to 1km in diameter then the core which powers it would be the size 
of a marble, or 1cm across. When we apply our simple models to such a 
complex system, it is worth remembering the enormous range of scales we 
are trying to understand. 

   The structure is that of a tiny C-O core surrounded by a He-burning shell 
and a H-burning shell, and then an enormous convective envelope. Following 
core He exhaustion the ignition of the He-burning shell is accompanied by a large  increase in the energy burning rate, at least for the more massive stars. This energy output causes the star to expand and the H-burning shell is extinguished. Regions progressively further inward become unstable to convection and thus the depth of the convective envelope increases, mixing the products of H burning to the surface in what is called the ``second dredge-up''. Note that this only occurs for masses above about 4 $\msun$, depending on the composition. In any event, during this so-called ``early AGB'' (E-AGB) phase the He shell forms and advances outward to meet the H shell. If second dredge-up occurs, it hastens the coming together of these two shells. 

  Soon after this the output of the He shell
becomes unstable, resulting in a thermal pulse or shell flash. Space 
prevents us from discussing the details of this, but basically it is 
caused by the high temperature dependence of 
the He-burning reactions as well as the
geometric thinness of the shell, which prevents if from responding to 
perturbations in the temperature. For details see \citet{HS72},  \citet{DS76}, 
\citet{S77}, \citet{D81}. Shell flashes are divided into four phases. 
During the ``on'' phase the He-shell provides the energy, and drives 
a convective zone from the He shell almost all the way to the H-shell (see 
Figure~1). This region contains about 75\% He and about 25\% C. 
During the ``power-down'' phase the convection dies away and the 
star expands in response to the energy input from the He burning. This 
expansion causes the outer regions to cool and the H shell is essentially 
extinguished. Just as happened during the second dredge-up, we find the 
bottom of the convective envelope moves inward (in mass) and can penetrate 
the region which was previously part of the intershell convective zone. 
This is called the ``dredge-up'' phase, and it mixes the freshly produced 
C to the surface. Finally the dredge-up stops, the star contracts again, 
the H shell is reignited and the star enters the ``off'' or ``interpulse'' 
phase. During this, the longest phase (about $10^4$ or $10^5$ y), the 
star is powered primarily by the H shell with negligible energy provided 
by the He shell.

\begin{figure*}   
  \centerline{\psfig{file=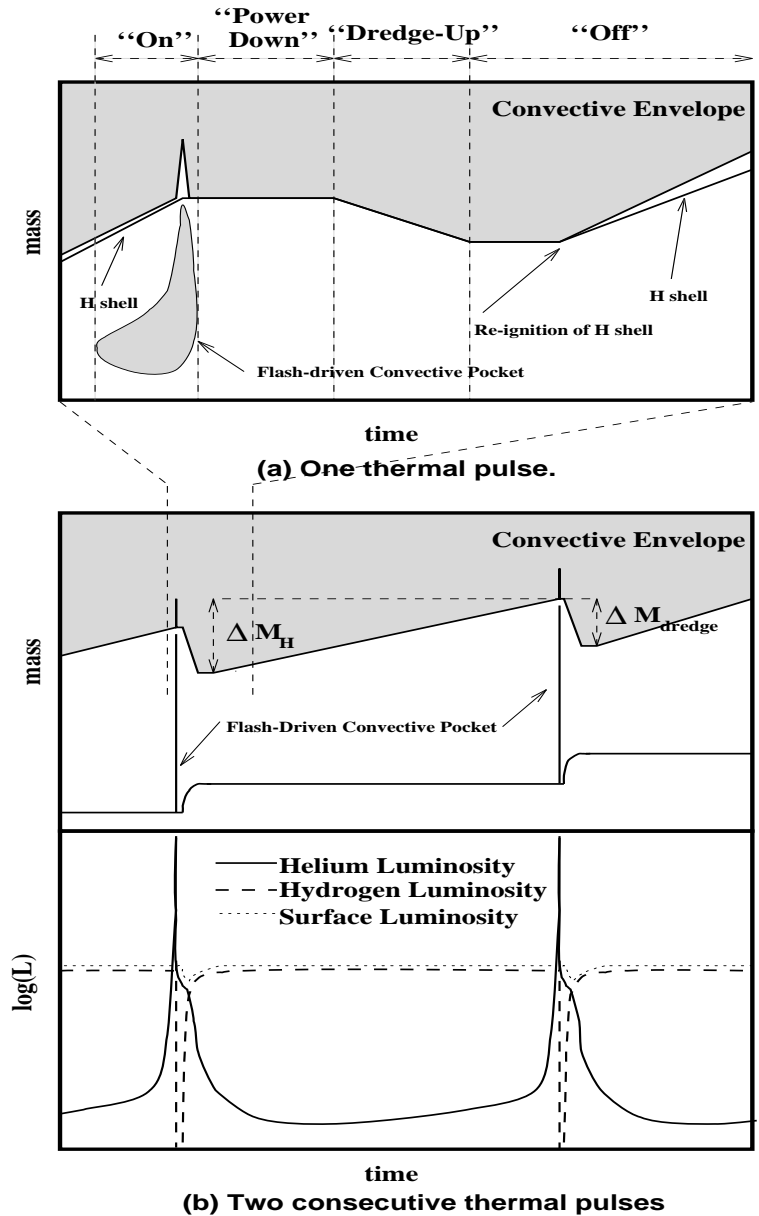,width=100mm,height=160mm,angle=0} }
  \caption{Schematic structure of an AGB star during the TP-AGB phase.}
  \label{fig:1} 
\end{figure*}

  These thermal pulses recur as the star ascends the AGB.
The evolution ends when the H shell runs out of fuel, i.e. when 
there is no longer an envelope for it to consume. Although the advance 
of the H shell is eating away at the bottom 
of the envelope, the dominant effect is mass-loss from the top of the 
envelope, as stressed by \citet{S79}. It is mass-loss which terminates 
the AGB evolution. But is it through steady loss? Or perhaps planetary 
nebula ejection? Or a combination? Does it depend on mass? Or on binarity? 
This is one of the major uncertainties (see below). We should note also 
that all AGB stars are long period variables of various kinds (eg Mira). 

%\subsection{Thermal Pulses}
We have seen how the thermal pulses produce C and hence can make carbon stars.
Observations show that AGB stars are usually enriched in s-process elements, 
and given the low temperatures seen in most of these (lower mass) stars, 
it seems that the neutron source is $^{13}$C$(\alpha,n)^{16}$O. The obvious
source of $^{13}$C is from CN cycling in the H shell, but this is not enough 
to produce the observed s-process enhancements. Pioneering work by 
\citet[][b]{IR82a} showed that some partial mixing can occur at the bottom 
of the H envelope during the dredge-up phase. If this happens, then some 
protons are deposited in the intershell region, which is about 25\% C. 
During the subsequent contraction, this $^{12}$C captures a proton to make 
$^{13}$C. If there are many protons then proton capture can continue and 
produce $^{14}$N (this is simply the CN cycle), so we do not want too many 
protons mixed into this region. In any event, there will now be a layer of 
enhanced $^{13}$C abundance,
referred to as a $^{13}$C pocket (see figure 2). During the subsequent 
interpulse phases this $^{13}$C produces neutron via $\alpha$-capture and 
these neutrons are captured by Fe (et al) and s-processing occurs in the 
$^{13}$C-pocket. During the next pulse this zone and its s-process elements 
is mixed into the convective shell. Also the higher temperatures at the 
bottom of the shell provide a brief burst of neutrons from 
$^{22}$Ne$(\alpha,n)^{25}$Mg. (The $^{22}$Ne has been produced by 
$\alpha$-captures on $^{14}$N which in turn was produced by the CNO cycle 
from the initial CNO.) Then the dredge-up phase mixes the resulting 
s-process elements to the surface of the star, where they are observed 
by spectroscopists. Many detailed calculations of this process have 
been published by the Torino group \citep[eg][]{G98}.

\begin{figure*}   
  \centerline{\psfig{file=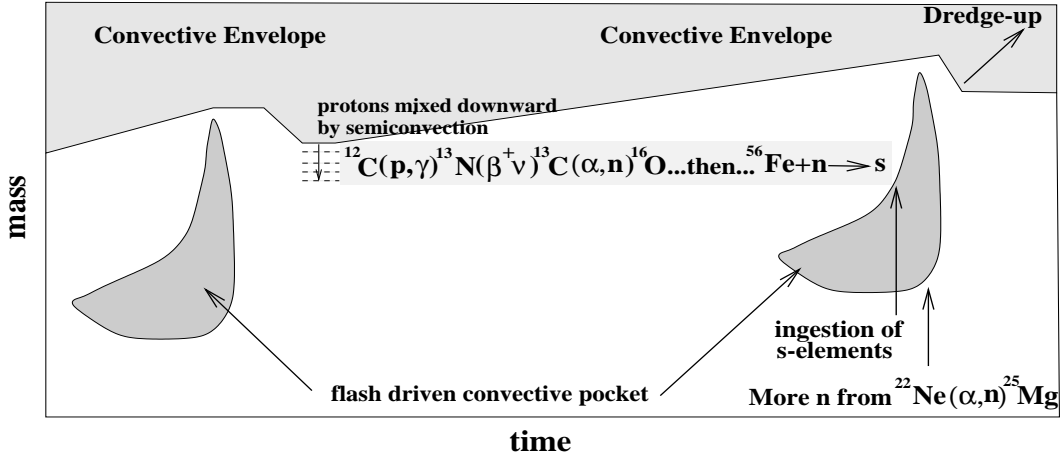,width=140mm,height=60mm,angle=0} }
  \caption{Schematic showing s-process production in two consecutive pulses.}
  \label{fig:2} 
\end{figure*}

Another important issue is the production of $^{19}$F, which is observed to 
correlate with the C/O ratio \citep{J92}. Theoretical studies show that 
F is indeed produced in pulses (space limitations prevent us from going 
into the details), but also requires an additional $^{13}$C source to 
explain the most abundant cases \citep[eg][]{MJA98}.

We have seen that $^{22}$Ne is abundant in the intershell. Hence thermal 
pulses can make $^{25}$Mg and $^{26}$Mg via $(\alpha,n)$ and  
$(\alpha,\gamma)$ respectively. Dredge-up then enriches the surface 
in these species. Similarly, the Ne-Na cycle can produce $^{23}$Na \citep{Mow}.

For more massive stars (above about $4\msun$, depending on 
composition) the bottom of the envelope extends into the top of the 
H burning shell. This means that some processing occurs in 
the envelope: this is called ``hot bottom burning'' (hereafter HBB). 
The first consequence of this is the production of $^7$Li via the 
Cameron-Fowler mechanism \citep{SB92}. Basically, mixing takes the 
$^7$Be (made by $^3$H+$^4$He) to cooler regions where it can produce 
$^7$Li rather than completing the PPII and PPIII chains. This matches 
observations of super-Li rich AGB stars in the Magellanic Clouds.
Of course, CN cycling will also occur when HBB exists \citep{BSA93}, 
so we expect $^{12}$C to be processed into $^{13}$C and $^{14}$N. 
Note that this can stop a star from becoming a carbon star, 
as the (primary) $^{12}$C mixed to the surface is 
then  burned into (primary) $^{13}$C and $^{14}$N. Hence another signature 
of HBB (as well as enriched $^7$Li) is an equilibrium ratio of the C isotopes.
We also see Ne-Na cycling and Mg-Al cycling, producing enrichments in 
$^{26}$Al, for example, although this element is also made in the H shell 
and enriched during dredge-up also \citep{MM}.

%Mg and Al
\section{From Giant to Giant: Aluminium in Globular Clusters}
An active area at present involves Mg and Al, which is 
relevant to this meeting because of the decay of $^{26}$Al to $^{26}$Mg.
It also involves AGB star nucleosynthesis and abundance anomalies
in present day globular cluster stars.

Globular clusters (GCs) are traditionally thought of as chemically homogeneous,
and this is true if one considers the Fe/H ratio for stars in a given 
cluster. The Fe content is pretty constant from star to star within a GC, although it varies substantially from one GC to another. At this stage we should note that most observations are done in the brightest giants, because of the difficulty of obtaining high resolution spectra for fainter stars. 

    But we soon discovered that other elements were less homogeneous. The 
first anomaly  was an anti-correlation of C with N---in stars where C was depleted, N was enhanced. This was soon extended to O and Na, which also anti-correlate. For reviews see \citet{K94} and \citet{Dac97}. These correlations clearly implicate hydrogen burning, and it was suggested by \citet{SM79} that the data could be understood as reflecting varying amounts of H-burning at the top of the hydrogen-shell. They proposed that some form of ``extra-mixing'', possibly meridional mixing,  may circulate the surface material down to the top of the H-shell where processing can occur, and then return the burnt material to the surface. Evidence for this was the clear decrease in C content with luminosity on the giant-branch for some clusters (such as NGC6397 and M92). More recently, the abundance anomalies in GC red-giants (GCRGs) have included a Mg-Al anti-correlation, as would be expected for the operation of the Mg-Al cycle. But the story gets very messy very quickly. 

   There are two major complications with the proposed mixing mechanism. Firstly, very recent observations with 8m-class telescopes have shown that some of these correlations exist  in unevolved stars in some clusters! eg 47~Tuc shows the C-N anti-correlation on the main-sequence \citep{C98} and NGC6752 shown the O-Na and Mg-Al anti-correlation at the base of the giant-branch and at the turn-off \citep{G01}. It is hard to see how this could arise from ``deep mixing''. Rather, this is evidence for the abundance anomalies being present in the gas from which the present-day GCRGs formed. In other words, this is evidence for what is called the ``primordial'' explanation. 

   The second complication comes from the only successful measurement of the abundances of the Mg isotopes, by \citet{Shet96}. He showed that in M13 there is an anti-correlation between $^{24}$Mg and Al, with the sum 
$^{25}$Mg$ + ^{26}$Mg being constant (Shetrone was unable to separate the two heavier Mg isotopes from each other, spectroscopically). He also found that as the Mg abundance decreased, it was $^{24}$Mg that was decreasing, and not the heavier isotopes! This implies that $^{24}$Mg is being burned into Al (presumably $^{27}$Al, because the evolution timescale for these stars exceeds $10^8$ years).
This is contrary to our understanding of the Mg-Al cycle and, if correct, poses a problem for theory.

   So the current situation is that some GCRGs show evidence for ``deep mixing'' bringing the products of H-burning to their surfaces (eg abundance variations correlating with evolutionary state on the giant branch), and others show evidence for primordial inhomogeneities (eg abundance variations in unevolved stars on the main-sequence). It appears that both processes occur, although the details probably depend on the mass and metallicity of the GC. It is worth pointing out that red-giants in the field show little, if any, of these effects! It is something peculiar to the GC environment.

  But how does this relate to AGB stars? Well, whatever stars polluted the GC must not have altered the Fe abundance, but they did alter the abundances of those things that take part in H-burning, such as O, Na, Mg and Al. It is thus very likely that the polluters were intermediate mass AGB stars \citep{Detal98}. Dredge-up will produce large amounts of the heavy Mg isotopes, and HBB will cycle these into both $^{26}$Al and  $^{27}$Al, and the  $^{26}$Al will then decay back to  $^{26}$Mg. Also, if one starts with an intermediate mass AGB star with a composition appropriate to a mixture of Big Bang material and supernova ejecta, then the initial Mg isotope ratios are about 98:1:1. Yet after some thermal pulsing the ratio goes closer to 50:25:25 due to the strong dredge-up of the heavy Mg isotopes. Further, \citet{benthesis} has shown that if the initial 
$^{24}$Mg  is underabundant compared to $^{25}$Mg and $^{26}$Mg then it becomes easier to match the observed Mg-Al correlations.

A recent development, and one of great interest to this meeting, is the suggestion by \citet{DW01} that the Al enhancements seen in the GCRGs are actually $^{26}$Al rather than $^{27}$Al. It appears than with the best reaction rates, it is simply impossible to make $^{27}$Al inthe required amounts. However the $^{26}$Al is more easily made, in larger amounts, and at lower temperatures, so that there is $^{26}$Al available for deep-mixing to transport to the surface. It is an intriguing idea, and may soon be open to an observational test.

Clearly the story is still unfolding, but it appears that earlier generations of  intermediate mass AGB stars, with highly non-solar compositions, have polluted the gas from which some GCRGs have formed. This may explain the abundance anomalies seen on the main-sequence. As these stars evolve, they also experience ``deep mixing'' which further alters their surface compositions. This combined scenario \citep{Detal98} explains many features, qualitatively, but needs to be investigated with quantitative models.

\section{Warnings and Uncertainties}
  It is usual to quote the basic input physics as uncertainties:~such
things as opacity and reaction rates. While it would be foolhardy to claim 
that the opacity is now known, it does appear that the recent work of the 
OPAL and OP groups have removed the biggest uncertainties. We do not think 
that opacity is a major uncertainty in AGB stars, but we do note that the 
resultant C-rich envelopes require appropriate opacity tables. 
Uncertainties in the reaction rates certainly remain a major problem, 
and we defer to \citet{Arnould} for a recent survey of this area.

Quantitatively the biggest uncertainties are in the areas of dredge-up 
and mass-loss. There is no good theory for mass-loss in very cool giants. 
One is forced to resort to empirical fits of observations. Although 
many formulae exist (see those listed in \citet{G} for example) none is 
clearly superior, although the Reimers formula  does not produce the 
very high  mass loss rates seen at the tip of the AGB.

Although it is common to bemoan the lack of a good convection theory, the main
limitation in modelling AGB stars is in determining the convective boundaries.
Some debate exists about  how to apply the Schwarzschild criterion (eg see \citet{Mowlavi2}) and the details of the numerical implementation \citep{FL}. These things greatly affect the depth of dredge-up, and limit the quantitative predictability of the models.

Related to this is the formation of the $^{13}$C pocket, which is clearly associated with the dredge-up problem. A partial mixing algorithm \citep{Falk} may provide a solution to this problem, but it contains some parameters which must be fixed somehow. Note that our current understanding has the $^{13}$C pocket produced by mass motions of some kind (possibly shear at the bottom of a rotating envelope, possibly overshooting, etc) and that this should not be very dependent on the composition. Hence the number of neutrons produced should be almost
independent of the metallicity of the star. So as [Fe/H] decreases, we expect a larger number of neutrons per seed, and a different s-process to occur in the very metal-poor stars. This does not seem to be the case \citep{ryan} and is another challenge for theorists!

\section{Summary}
An AGB star is a complicated thing, but that is what makes it interesting.
The interplay of different kinds of physics is fascinating. Reliable 
quantitative estimates of their behaviour require addressing the areas of uncertainty we mention above. Before placing too much faith in these estimates, look carefully to see which assumptions went into the details-there lies the devil, as always.

% Bibliographic references with the natbib package:
% Parenthetical: \citep{Bai92} produces (Bailyn 1992).
% Textual: \citet{Bai95} produces Bailyn et al. (1995).
% An affix and part of a reference:
%   \citep[e.g.][Ch. 2]{Bar76}
%   produces (e.g. Barnes et al. 1976, Ch. 2).


\begin{thebibliography}{}
\bibitem[Aoki et al(2000)]{ryan}Aoki, W., et al, 2000, Ap J, 536, L97.
\bibitem[Arnould et al(1999)]{Arnould}Arnould, M., Goriely, S., and Jorissen, A., 1999, Astron Astrophys, 347,
 572
\bibitem[Boothroyd et al(1993)]{BSA93}Boothroyd, A. I., Sackmann, I.-J., and Ahearn, S. C., 1993, Ap J, 416, 762.
\bibitem[Cannon et al(1998)]{C98}Cannon, R. D., etal, 1998, MNRAS, 208, 601.
\bibitem[Da~Costa(1997)]{Dac97}Da~Costa, G. S., 1997, in ``Fundamental Properties of Stars'', eds T.R. Bedding et al (Dordrect: Kluwer).
\bibitem[Denissenkov et al(1998)]{Detal98}Denissenkov, P., et al, 1998, Astron Astrophys, 333, 926.
\bibitem[Denissenkov \& Weiss(2001)]{DW01}Denissenkov, P., and Weiss, A., 2001, Ap J Lett, 559, 115.
\bibitem[Despain(1981)]{D81}Despain, K. H., 1981, Ap J, 251, 639.
\bibitem[Despain \& Scalo(1976)]{DS76}Despain, K. H. and Scalo, J.M., 1976, Ap J, 208, 789.
\bibitem[Frost \& Lattanzio(1996)]{FL}Frost, C. A., and Lattanzio, J. C., 1996, Ap J, 473, 383.
\bibitem[Gallino et al(1998)]{G98}Gallino, R., et al, Ap J, 497, 388.
\bibitem[Gratton, et al(2001)]{G01}Gratton, R. G., et al, 2001, Astron Astrophys, 369, 87.
\bibitem[Groenewegen \& de Jong(1994)]{G}Groenewegen, M. A. T.., and de Jong, T. , 1994, Astron Astrophys, 283, 463.
\bibitem[H\"arm \& Schwarzschild(1972)]{HS72}H\"arm, R. and Schwarzschild, M., 1972, Ap J, 172, 403.
\bibitem[Herwig(2000)]{Falk}Herwig, F., 2000, Astron Astrophys, 360, 952.
\bibitem[Iben \& Renzini(1982a)]{IR82a}Iben, I., Jr., and Renzini, A., 1982a, Ap J Lett, 259, L79.
\bibitem[Iben \& Renzini(1982b)]{IR82b}Iben, I., Jr., and Renzini, A., 1982b, Ap J Lett, 263, L231.
\bibitem[Jorissen et al(1992)]{J92}Jorissen, A., et al, 1992, Astron Astrophys, 261, 164.
\bibitem[Kraft(1994)]{K94}Kraft, R. P., 1994, Pub ASP, 106, 553.
\bibitem[Messenger(2000)]{benthesis}Messenger, B. B., 2000, PhD Thesis, Monash University.
\bibitem[Mowlavi(1999a)]{Mow}Mowlavi, N., 1999a, Astron Astrophys, 350, 73.
\bibitem[Mowlavi(1999b)]{Mowlavi2}Mowlavi, N., 1999b, Astron Astrophys, 344, 617.
\bibitem[Mowlavi et al(1998)]{MJA98}Mowlavi, N., Jorissen, A., and Arnould, M., 1998, Astron Astrophys, 334, 153.
\bibitem[Mowlavi \& Meynet(2000)]{MM}Mowlavi, N., and Meynet M., 2000, Astron Astrophys, 361, 959.
\bibitem[Sackmann(1977)]{S77}Sackmann, I.-J., 1977, Ap J, 212, 159.
\bibitem[Sackmann \& Boothroyd(1992)]{SB92}Sackmann, I.-J., and Boothroyd, A. I., 1992, Ap J, 392, L71.
\bibitem[Sch\"onberner(1979)]{S79}Sch\"onberner, D., 1979, Astron Astrophys, 79, 108.
\bibitem[Shetrone(1996)]{Shet96}Shetrone, M., 1996, Astron J, 112, 2639.
\bibitem[Sweigart \& Mengel(1979)]{SM79}Sweigart, A. V., and Mengel, J., 1979, Ap J, 229, 624.
% \bibitem[Names(Year)]{label} or \bibitem[Names(Year)Long names]{label}.
% (\harvarditem{Name}{Year}{label} is also supported.)

%\bibitem[]{}

\end{thebibliography}
\end{document}